\documentclass[aps,prb,twocolumn,superscriptaddress,showpacs]{revtex4-1}
\usepackage{hyperref}
\usepackage{graphicx}
\usepackage{revtex4-1}

\newcommand{\SFA}{SrFe$_{2}$As$_{2}$}
\newcommand{\CFA}{CaFe$_{2}$As$_{2}$}

\newcommand{\Sqw}{S$(\mathbf{q},\omega)$} 
\newcommand{\qvec}{$\mathbf{q}$}

\newcommand{\degree}[1]{$#1^\circ$}
\newcommand{\appr}{$\approx$}

\begin{document}

\title{Phonon spectrum of SrFe$_{2}$As$_{2}$ determined using multizone phonon refinement}

\author{D.~Parshall}
	\affiliation{University of Colorado, Department of Physics, Boulder, CO 80309}
	\altaffiliation{NIST Center for Neutron Research, National Institute of Standards and Technology, Gaithersburg, Maryland 20899, USA}
	\email{parshall@nist.gov}

\author{R.~Heid}
	\affiliation{Karlsruhe Institut f\"ur Technologie, Institut f\"ur Festk\"orperphysik, P.O.B. 3640, D-76021 Karlsruhe, Germany}

\author{J.~L.~Niedziela}
	\affiliation{Instrument and Source Division, Oak Ridge National Laboratory, Oak Ridge, TN 37831}

\author{Th.~Wolf}
	\affiliation{Karlsruhe Institut f\"ur Technologie, Institut f\"ur Festk\"orperphysik, P.O.B. 3640, D-76021 Karlsruhe, Germany}

\author{M.~B.~Stone}
	\affiliation{Quantum Condensed Matter Division, Oak Ridge National Laboratory, Oak Ridge, TN 37831}

\author{D.~L.~Abernathy}
	\affiliation{Quantum Condensed Matter Division, Oak Ridge National Laboratory, Oak Ridge, TN 37831}

\author{D.~Reznik}
	\affiliation{University of Colorado, Department of Physics, Boulder, CO 80309}

\date{\today}

\begin{abstract}The ferropnictide superconductors exhibit a sensitive interplay between the lattice and magnetic degrees of freedom, including a number of phonon modes that are much softer than predicted by nonmagnetic calculations using density functional theory (DFT).  However, it is not known what effect, if any, the long-range magnetic order has on phonon frequencies above 23~meV, where several phonon branches are very closely spaced in energy and it is challenging to isolate them from each other. We measured these phonons using inelastic time-of-flight neutron scattering in $\approx$ 40~Brillouin zones, and developed a technique to determine their frequencies.  We find this method capable of determining phonon energies to $\approx$~0.1~meV accuracy, and that the DFT calculations using the experimental structure yield qualitatively correct energies and eigenvectors.  We do not find any effect of the magnetic transition on these phonons.
\end{abstract}


\maketitle

\section{Introduction}

The role of phonons in the mechanism of high-temperature superconductivity in ferropnictides remains poorly understood.  While density function theory (DFT) calculations indicate that conventional electron-phonon coupling is not responsible for superconductivity \cite{Boeri08}, there is nevertheless a strong relationship between the lattice and electronic properties, including both magnetism and superconductivity.   For example, there is an isotope effect due to Fe \cite{Liu09}, and the superconducting temperature strongly depends upon the Fe-As-Fe bond angle \cite{Lee08} (or alternatively the height of the pnictogen atom above the plane of the transition metal\cite{Mizuguchi10}).

In addition, the lattice and magnetism are strongly coupled, which manifests in several ways.  First, in order to correctly refine the internal parameters of the structure, DFT calculations must be performed with spin-polarization \cite{Mazin09}.  Second, \CFA~enters a ``collapsed'' tetragonal state at mild isotropic pressures \cite{Kreyssig08}, in which the c-axis contracts by \appr 10\% and the magnetism becomes quenched \cite{Goldman09}; this is the result of strong coupling between the Fe moment and interplanar As-As bonding \cite{Yildirim09}.   Third, the increase of the static moment with As-Fe separation can be described by a Landau/Stoner theory \cite{Egami10}, in which a larger volume around each Fe atom allows a larger magnetic moment.  This dependence of the moment on As position has been confirmed experimentally \cite{Egami10-1} for the static moment in the CeFeAs$_{1-x}$P$_x$O series.

This strong magnetostructural coupling can in turn influence the phonon spectrum.  For instance, the frequency of the Raman-active mode corresponding to vibration of the As atoms is not predicted correctly unless the Fe moment is taken into account \cite{Reznik09, Hahn09}.  In addition, the phonon density of states in LaFeAsO is appreciably softer than predicted by nonmagnetic DFT \cite{Fukuda08}, especially for the higher-energy modes corresponding to in-plane motion of the Fe and As atoms.  While the dispersion of the c-axis modes has been studied elsewhere \cite{Reznik09, Hahn09}), we are not aware of any reports regarding the in-plane Fe and As phonons.

Here we present measurements of the phonon dispersion in \SFA~obtained through time-of-flight (TOF) inelastic neutron scattering, focusing on the closely-spaced higher-energy modes which are sensitive to Fe-As coupling and compare these results to the DFT calculations. In order to isolate the dispersions of different phonons from nearby branches, we developed a novel approach that utilizes the entire TOF spectrum across many Brillouin zones (BZ), which we call multizone phonon refinement (MPR).  We show that the experimental phonon dispersions agree well with DFT calculations performed with the structure constrained to the measured structure. We also found that there is no significant temperature dependence across the magnetic ordering transition of any of the phonons for which it was measured.


\section{Experimental details}
\begin{figure}[br]
   \centering
   \includegraphics[scale=0.27]{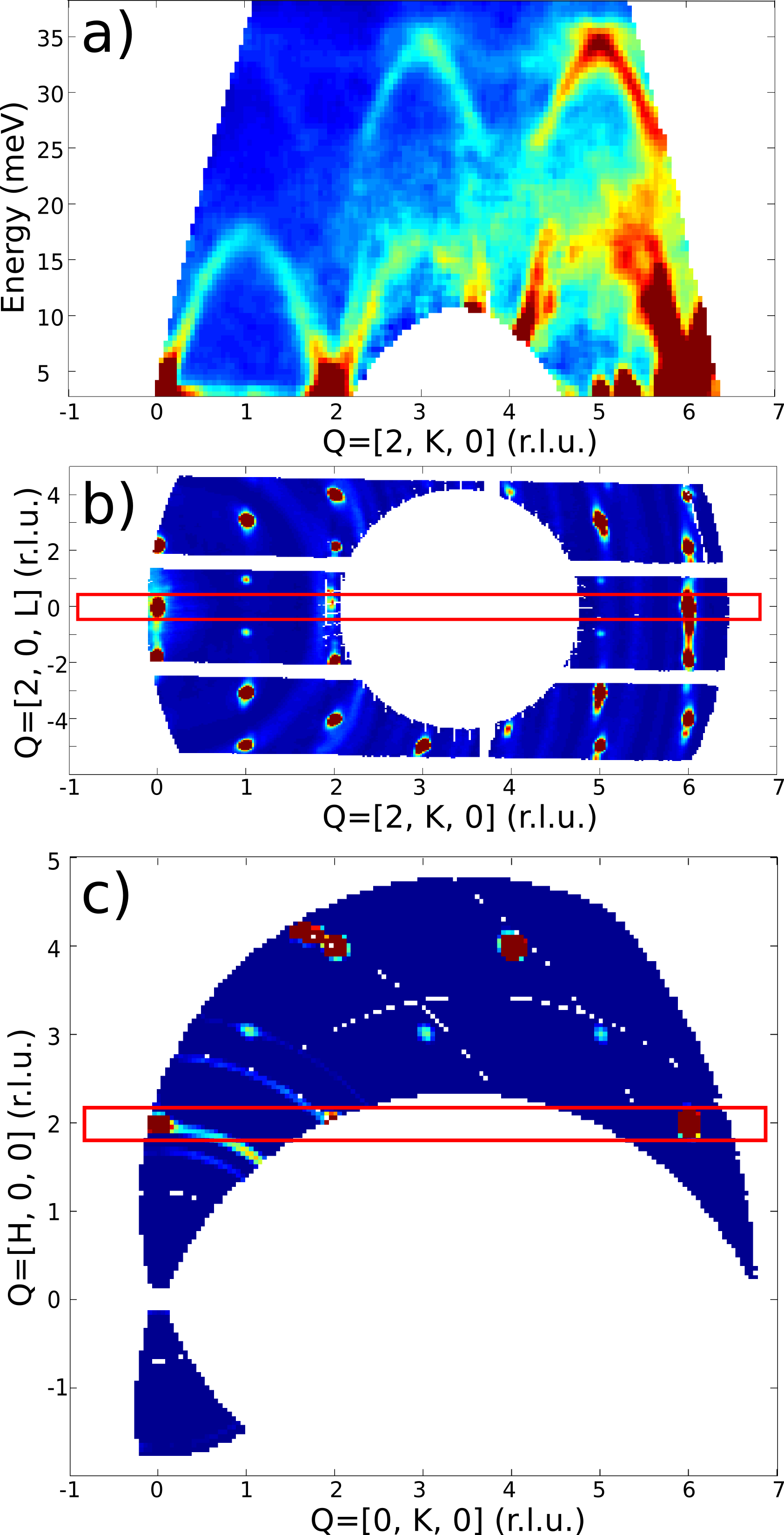} 
    \caption{(color online) Reciprocal space mapped out in the experiment at T~= 10~K.   For all panels, the horizontal axis plots K from -1 to 7.  $\mathbf{a)}$ Phonon dispersion along (2,K,0).  This is the inelastic scattering for the regions shown as red boxes in panels $\mathbf{b}$ and $\mathbf{c}$.  Near (2,0,0) the phonons are strictly transverse, but near (2,6,0) the longitudinal phonons are also evident.  $\mathbf{b)}$  Map of elastic scattering in the (2,K,L) plane, showing many Bragg peaks.  $\mathbf{c)}$  Map of elastic scattering in the (H,K,0) plane, showing 8 Bragg peaks.  
}
   \label{fig:slices}
\end{figure}

The experiment was performed using the ARCS time-of-flight spectrometer at the Spallation Neutron Source, located at the Oak Ridge National Laboratory \cite{Abernathy2012}.  Data were collected using the recently-developed rotating sample technique \cite{Weber2012, Horace}, which allows us to collect a 4-dimensional dataset (H, K, L, energy).  

The sample was a single crystal of \SFA, grown from self-flux in glassy carbon crucibles \cite{Hardy2010}, using extra-low cooling rates of \degree{0.22} - \degree{0.25} C/h.  The sample mass was \appr~1.5~g, and the mosaic was \appr~\degree{2}.  At room temperature the material is tetragonal (space group 139, \emph{I4/mmm}), and at \appr~210~K undergoes an AFM/orthorhombic transition \cite{Kaneko08} (space group 69, \emph{Fmmm}).  Because the orthorhombic distortion is quite small ($2(a-b)/(a+b)$~\appr~1\%), we use tetragonal notation throughout this paper.  The sample was mounted using twisted aluminum foil, and placed in a closed-cycle refrigerator.  No exchange gas was used, so the temperatures are accurate to within 10~K.

Data were collected over a period of 6 days; approximately 3 days over an angular range of \degree{39} at T~= 10~K, and approximately 1.5 days over a range of \degree{15} at both T~= 220~K and T~= 320~K.  This corresponds to roughly 45 BZ in the energy range of interest for 10~K and 20 BZ for 220~K (just slighly above the magnetic ordering temperature), and 320~K.  The sample was mounted in the HK0 plane, but the out-of-plane detector coverage allows access up to L \appr~$\pm$~4.  The incident energy was 70~meV, with a FWHM of \appr~3~meV at the elastic line, and \appr~1.5~meV in the range of interest.

The details of the DFT calculations have been described previously \cite{Heid1999}.  We used the non-magnetic tetragonal unit cell, with the experimental lattice parameters and atomic positions taken from previously-published work \cite{Tegel08}.  The eigenvectors are closely related to the atomic displacements, which in turn determine the intensities of the phonon peaks in various BZ.  These eigenvectors can be calculated quite well at the high-symmetry points, and the calculation allows for the phonon intensity at other points to be calculated using a perturbative method.  These energies and intensities are then used as a starting point for the refinement, as described below.

\section{Multizone phonon refinement (MPR)}
\begin{figure}[t]
   \centering
   \includegraphics[scale=0.48]{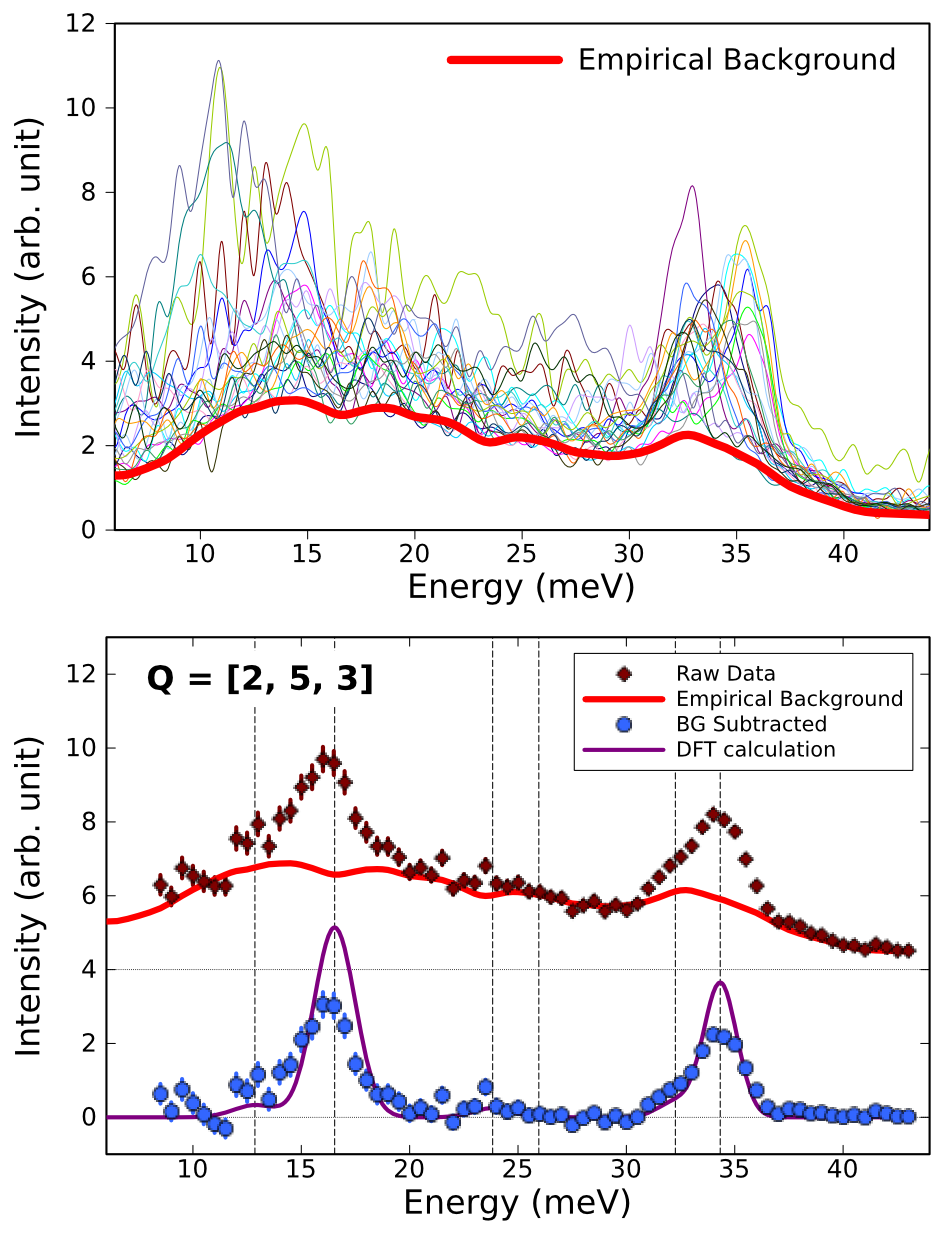} 
	\caption{(color online) Background determination.  \emph{Upper~panel:} the background (solid red line) is determined empirically by averaging the lowest intensities at each energy.  Shown here are constant-Q scans from 25 arbitrary points in reciprocal space.  \emph{Lower~panel:}  Constant-Q data at $\mathbf{Q}~= [2,~5,~3]$ (gamma point).  The background (solid red line) can be subtracted from the raw data (red diamonds) to produce a remarkably clean spectrum (blue circles).  Dashed lines show the refined values of the peak centers.  The intensities shown are predicted from DFT, which provides reasonable initial conditions for least-squares fitting.
}
   \label{fig:BG_subt}
\end{figure}

\begin{figure}[b]
   \centering
   \includegraphics[scale=0.4]{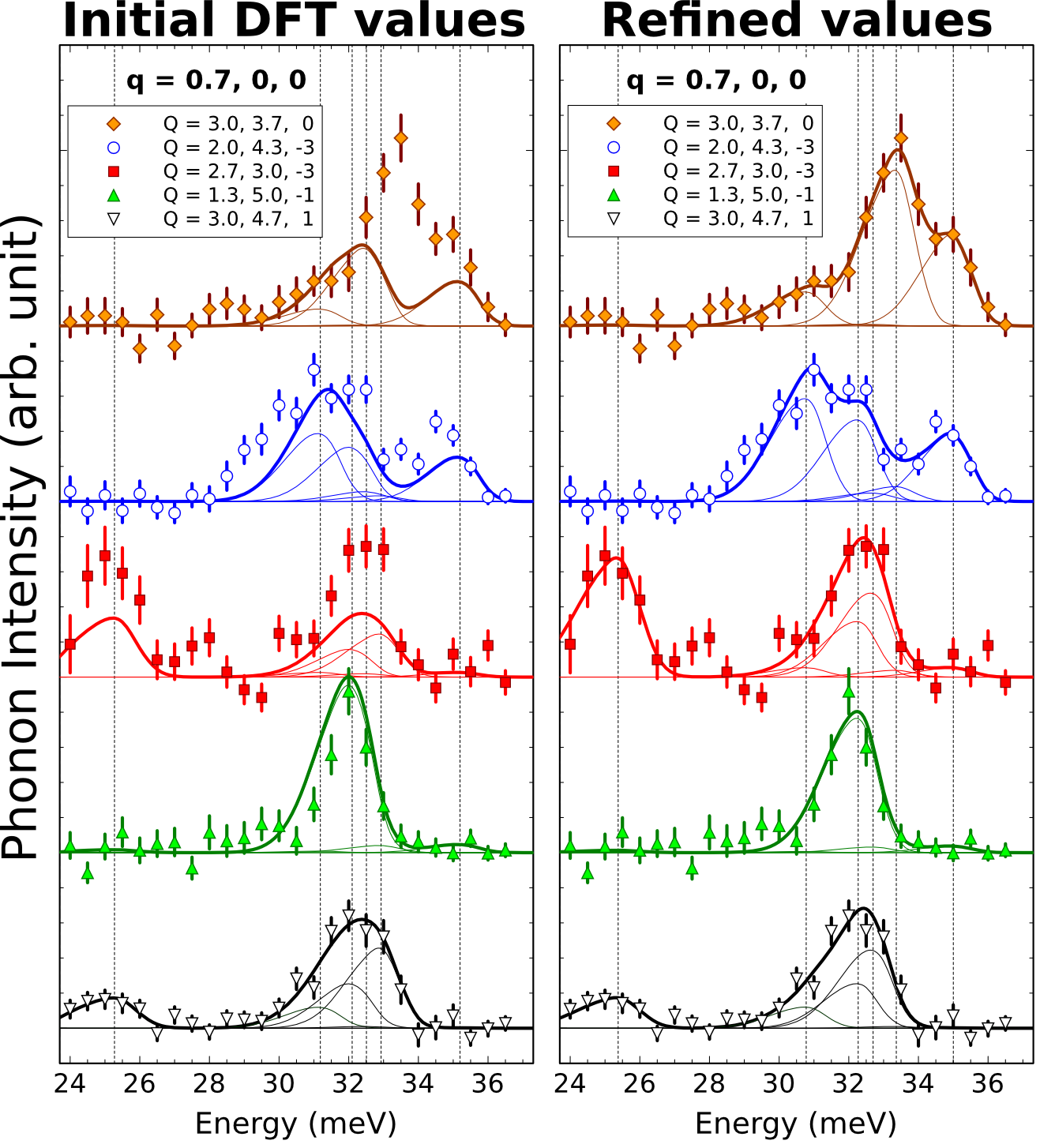} 
    \caption{(color online)  Data and model at \qvec = [0.7, 0, 0] from selected Brillouin zones.  The left panel displays the data (points with error bars) together with phonon energies and intensities as predicted by DFT; the right panel displays the same data, along with the energies and intensities after multizone phonon refinement. Although data from only 5 symmetry points are shown here, a total of 154 symmetry points were used for the fitting.  Data from different $\mathbf{Q}$ are vertically offset for clarity.  The model consists of 6 phonons each constrained to have the same energy and lineshape in all BZ.  The individual phonons broadened by experimental resolution are shown with thin lines, and the summation with solid lines.  The dashed vertical lines mark the phonon energies, and facilitate comparison between the different $\mathbf{Q}$.}
   \label{fig:multizone}
\end{figure}

\begin{figure*}[t]
   \centering
   \includegraphics[scale=0.45]{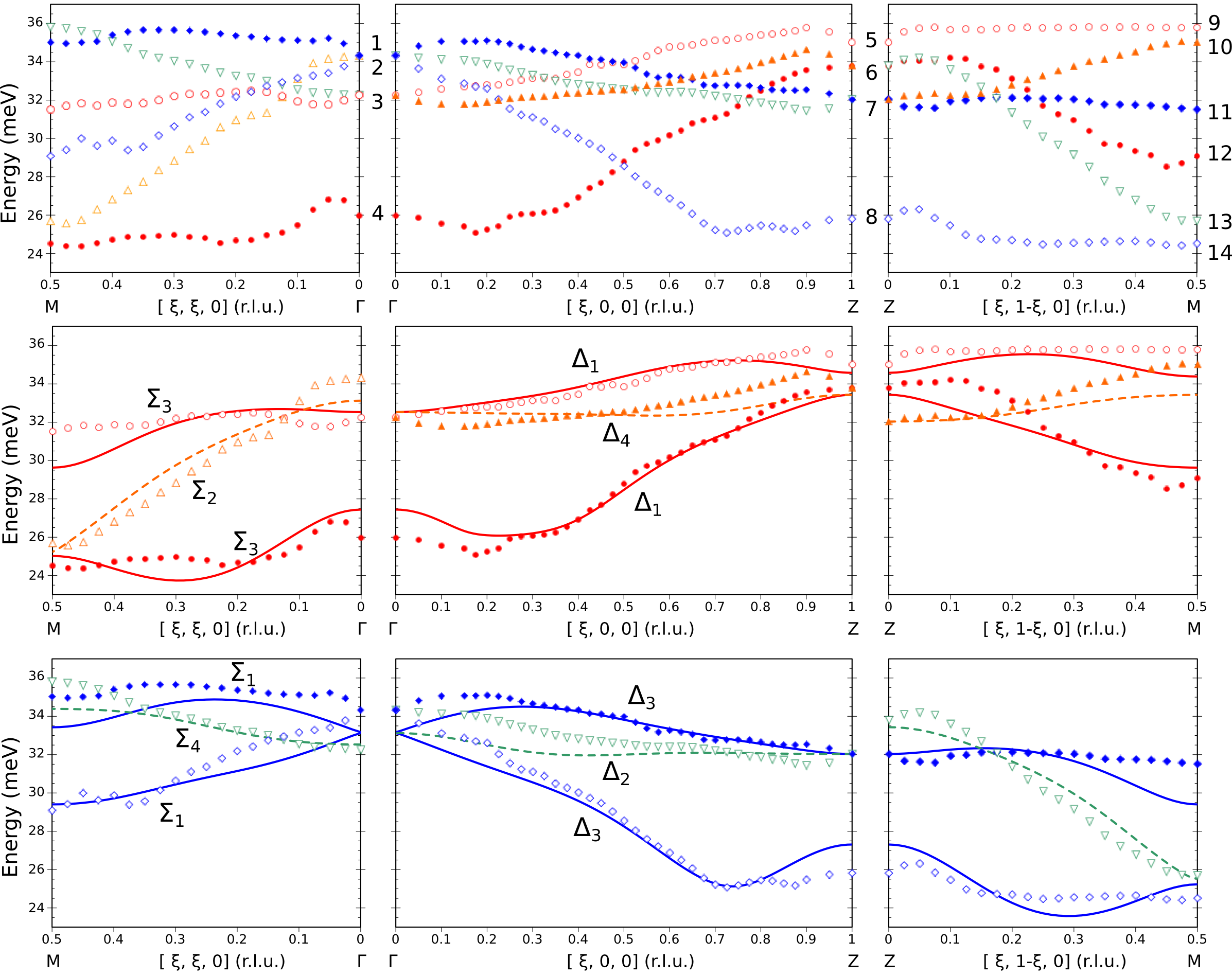} 
    \caption{(color online) Experimental dispersions at T~= 10~K obtained by MPR, and comparison with DFT calculations.  Along each direction (M-$\Gamma$, $\Gamma$-Z, Z-M), modes with the same symmetry are shown with the same color and shape, and the symmetry class is indicated.  Data have been smoothed by averaging with nearest-neighbor points.  The errorbars are not shown, but are generally approximately the size of the points.  \emph{Top panels:}  Measured dispersion of high-energy Fe-As modes in \SFA at T~= 10~K, running from M (far left), to $\Gamma$, to Z, to M (far right).  The numbers between each panel refer to the eigenvectors (shown in Fig. \ref{fig:eigs}).  \emph{Center and lower panels:} The same data as in upper panels, but just 3 modes per panel are shown, to facilitate comparison with DFT calculation (lines).  DFT energies have been increased by 5\%.}
   \label{fig:disp}
\end{figure*}

Optical phonons in many compounds with a large unit cell form dense ``spaghetti'' of branches in large parts of reciprocal space. In such a case energies of different phonons at the reduced wavevector are separated by much less than the energy-resolution of the spectrometer.  Quite often it is easy to mistake two nearby peaks for a single broad peak.  Here we outline a novel technique, multizone phonon refinement (MPR) that allows accurate determination of phonon dispersions in such cases.  We use this technique to resolve a group of branches between 24 and 36 meV in \SFA,~measured as described in the previous section.

\subsection{Background determination} 
In neutron-scattering experiments the inelastic background arises from a combination of factors, primarily incoherent phonon scattering by the sample, powder-averaged phonon scattering from the mount (in this case aluminum foil), secondary scattering (a Bragg peak and a phonon), and multiphonon scattering (strongest at high temperature).  The first factor necessarily produces isotropic scattering with a $\mathbf{Q}^2$ dependence characteristic of phonons. For this work all contributions to the inelastic background are assumed to follow this form.

In order to determine the background, we divided measured reciprocal space into equally-sized volume elements (known as ``voxels''), normalizing by $\mathbf{Q}^2$.  Each voxel corresponds to an energy scan at a fixed  $\mathbf{Q}$.  For a given energy, the mean of the lowest-intensity voxels was taken to be the background (see Fig.~\ref{fig:BG_subt}).  In other words, for each energy we found the lowest intensities over the entire measured dataset (after dividing by $\mathbf{Q}^2$) and took that to be the background.

Figure \ref{fig:BG_subt} shows such energy spectra plotted together in thin lines. The thick solid line corresponding to the average intensity of the lowest 1\% of the $\mathbf{Q}$-voxels.  We experimented with using the lowest 1\% to the lowest 5\% of the $\mathbf{Q}$-voxels at each energy, and found only minor changes to the resulting background function.  Since our primary interest here was the higher-energy phonons, we made no attempt to subtract the elastic line.  In future experiments we can subtract the elastic line separately from each voxel of $\mathbf{Q}$, and then subtract the background.  An additional linear term was included in the fitting to account for any residual intensity, but was generally insignificant. At lower $\mathbf{Q}$ there is some additional background, but these data have little effect on the overall fit. As can be seen in Fig.~\ref{fig:BG_subt}, the subtraction works quite well in the sense that it results in spectra that consist of phonon peaks on top of a very small linear background.

Note that because this method measures the background empirically, the source of the background (sample, aluminum, multiphonon, elastic line, etc.) is irrelevant; the only stipulation is that the angular range of data collection be large enough so that at some point of the measured range of \Sqw, the phonon structure factor goes to zero.  The precise range of $\mathbf{Q}$ required to meet that condition varies depending on symmetry of the sample and the energy range of interest.  In the worst-case scenario (collecting over the entire energy range), the \emph{I4/mmm} symmetry of this sample would require a rotation of \degree{45}, plus an additional range of \appr~\degree{15} to account for the different region of reciprocal space which is measured at different energy transfers.  In practice the actual range covered may be substantially less; we found no appreciable difference whether gathering the background from a \degree{39} or \degree{15} range.

One potential limit of the background subtraction procedure is in the case of a non-dispersive mode.  If a mode maintained both a constant energy and a nonzero structure factor, then some of its intensity might be incorrectly assigned to the background.  However, this situation can only occur for an isolated atom in an isotropic environment (a rattling mode inside of a rigid cage might approximate this condition).  Other than this special case, the intensity of a well-behaved mode must vary with $\mathbf{Q}$, due to either the phase factor, or to the projection of the eigenvector onto $\mathbf{Q}$.

\subsection{Multizone fitting} 
Traditionally, phonon measurements have been made at specific wavevectors in one or two BZ. This works when phonons are well-separated in energy, but in cases of multiple overlapping peaks it is often impossible to distinguish a single broad peak from two or more narrow peaks. TOF inelastic neutron scattering instruments capture data over many BZ. Thus phonon peaks belonging to a multiplet of overlapping peaks at a specific reduced wavevector will have different intensities in different BZ, but they will have the same energies and intrinsic linewidths. To perform the fitting, we first start with predictions of DFT calculations, which give approximate phonon frequencies and structure factors. Some of these are very accurate and agree with the data without any fitting, whereas others are significantly off. (see left panel in Fig.~\ref{fig:multizone}) By simultaneously fitting the experimental phonon spectrum at the same reduced wavevector in multiple BZ, we can identify experimental positions of individual phonons them quite well if we constrain their widths and energies to be BZ-independent while letting their intensities vary from zone to zone (see right panel in Fig.~\ref{fig:multizone}).

The energy resolution of ARCS is nearly independent of $\mathbf{Q}$ over the entire instrument, so values for resolution widths were calculated as a function of energy, as described elsewhere \cite{Abernathy2012}.  Because ARCS has an asymetric resolution function, the peaks were modeled as split Gaussians, with the lower (higher) side of the peak having 60\% (40\%) of the FWHM.  The peak width and asymmetry were checked at several $\mathbf{Q}$ for which DFT predicted only a single phonon was allowed in the 32-35 meV range, and this modeling of the energy resolution was found to be satisfactory.

While on a triple-axis instrument the momentum and energy resolutions are strongly coupled, for ARCS they are largely independent.  The $\mathbf{Q}$-resolution at higher momentum transfer (where phonons are strongest) is determined primarily by the sample mosaic.  Our sample mosaic of \degree{2} sets $\Delta_{\mathbf{Q}}$/$\mathbf{Q}$ \appr 3\%, on the same order as the $\mathbf{Q}$-resolution set by size of the detector pixels (\appr 1\%).  The in-plane voxel size (Q-binning) was either $\pm$ .025 r.l.u., or $\pm$ 0.05 r.l.u., which makes $\Delta_{\mathbf{Q}}$/$\mathbf{Q}$ between 1-2\% or 2-4\%, respectively.  The out-of-plane voxel size was $\pm$ 0.5 r.l.u. in all cases; this puts $\Delta_{\mathbf{Q}}$/$\mathbf{Q}$ \appr 6\%, but the DFT calculation indicates that the energy should vary by less than 1\% over this range (c-axis modes are generally fairly non-dispersive in layered systems such as this one), reducing the effect of $\mathbf{Q}$-resolution on determination of the phonon energies.

In order to simulate the combined effect of dispersion and finite bin size on the energy resolution, we increased the calculated energy FWHM by 10\%.  For weakly (strongly) dispersive branches, this will cause the refined separation between overlapping modes to be smaller (larger) than the true separation.  A more detailed analysis could modify the energy broadening based upon bin size and local dispersion as calculated from DFT. Other than this broadening, no attempt was made to account for the effect of $\mathbf{Q}$-resolution on the energy linewidth.

The fitting program was written in Matlab, using a standard least-squares algorithm.  Each reduced-$\mathbf{q}$ symmetry point was fit separately.  The heights of the peaks were constrained to be positive, but could vary between BZ.  Peak centers were constrained to be identical in various BZ.  Resolution widths were fixed as described above.  For each voxel, the empirical background was subtracted, and an additional linear term was included in the fitting to account for any residual background (but was generally small).  The energy linewidth was fixed to be resolution-limited, but can be made a fitted parameter in the future.


\section{Results and discussion}
\begin{figure}[t]
   \centering
   \includegraphics[scale=0.14]{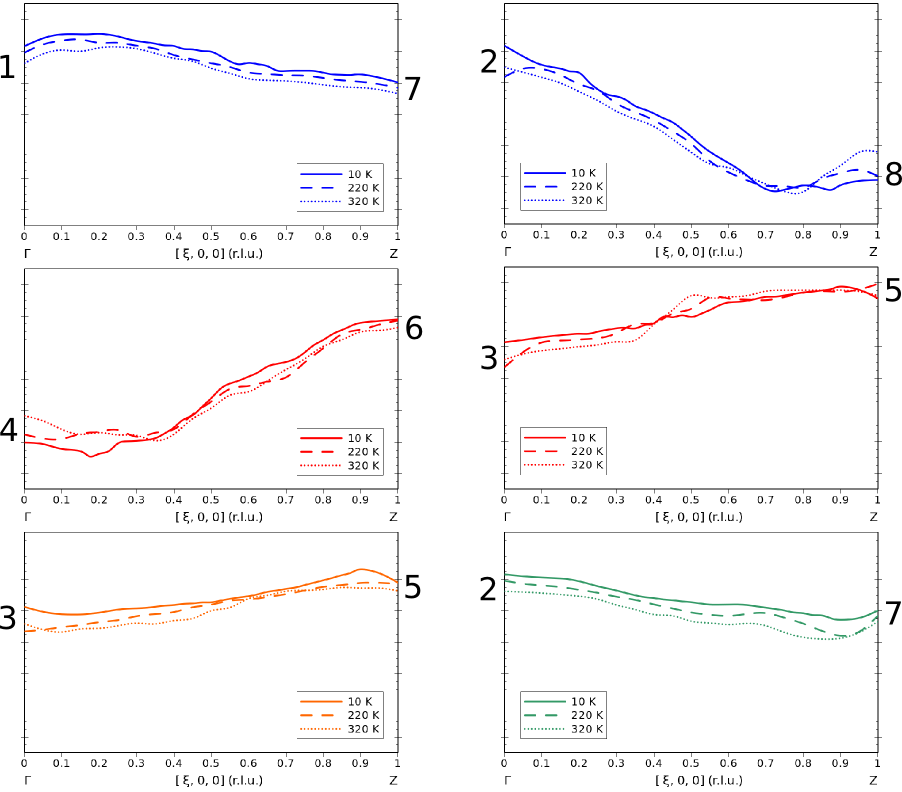} 
    \caption{(color online) Temperature dependence of each phonon branch along the $\Gamma$-Z line. Solid lines are from T~= 10~K, dashed lines from T~= 220~K, and dotted lines from T~= 320~K.  The data from T~= 220~K and 320~K were taken over a shorter time and smaller angular region and are correspondingly noisier.  The endpoints are labelled with the same numbers used in Fig. \ref{fig:disp}.}
   \label{fig:temp}
\end{figure}

\begin{figure*}[t]
   \centering
   \includegraphics[scale=0.1]{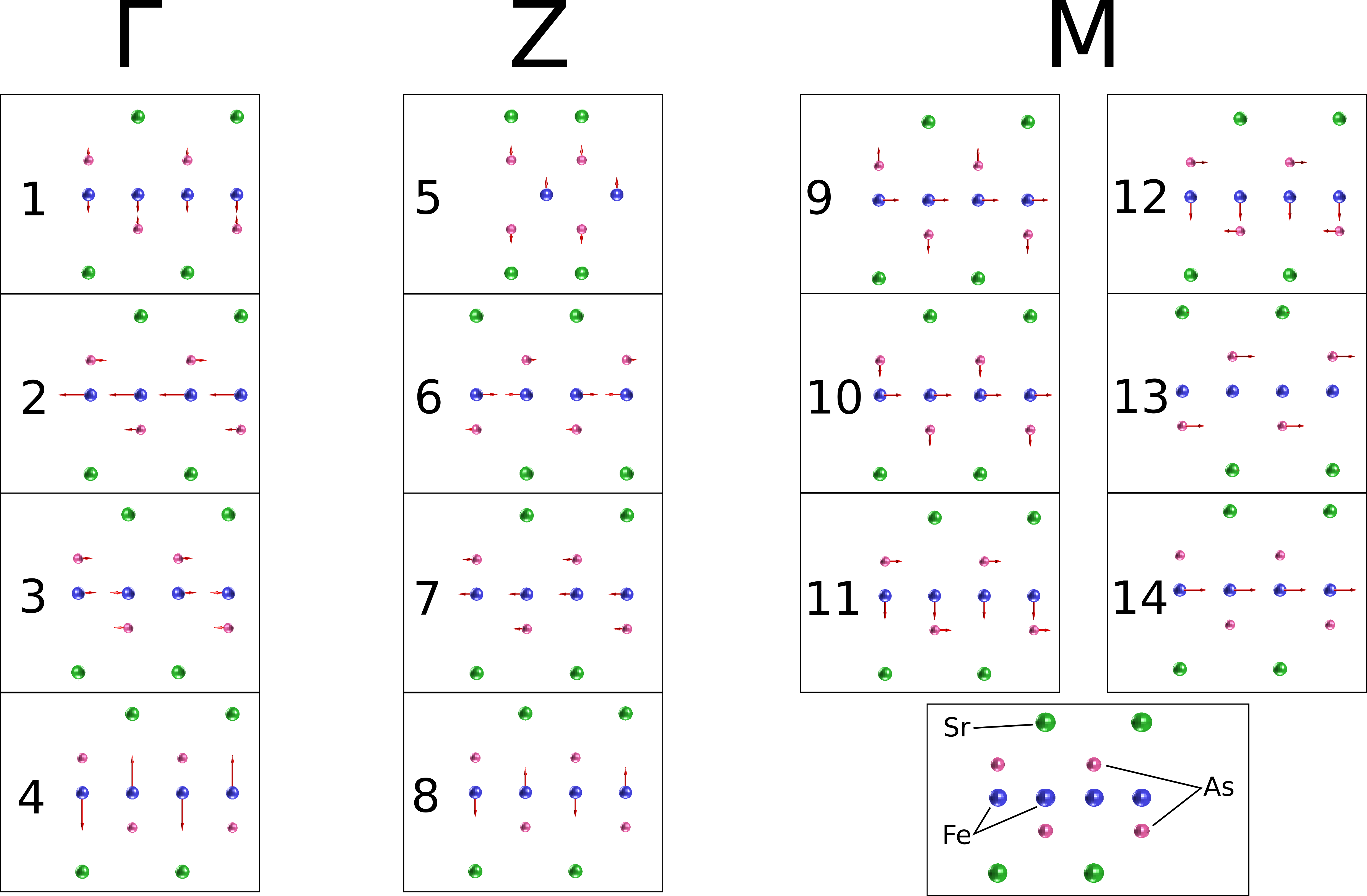} 
    \caption{(color online) Eigenvectors for phonons at high-symmetry points; the numbers correspond to Fig. \ref{fig:disp}. View is along the [100] axis; larger blue spheres are Fe, smaller pink spheres are As (see inset, lower right).  The arrows reflect the qualitative motion of the atoms, but do not attempt to represent the relative phase.  The in-plane motion of atoms along the $\Gamma$-Z line is primarily along the [100]/[010] direction.  The in-plane motion along  $\Gamma$ and M point is primarily along the [110]/[-110] direction (i.e., \degree{45} from the viewing axis), but this viewing orientation makes it clear which rows of Fe atoms are moving in unison.}
   \label{fig:eigs}
\end{figure*}

Figure \ref{fig:disp} shows extracted dispersions and branch assignments for narrowly-spaced phonon branches in \SFA.  Note that their separation in energy was often much smaller than the experimental resolution.  Our method of determining the phonon peak positions automatically provides BZ-dependence of peak intensities for each reduced $\mathbf{q}$, which is determined by the phonon eigenvectors.  Thus we assign phonon peaks to the same branch if their eigenvectors vary smoothly throughout the zone (a similar method has been previously used to assign eigenvectors using triple-axis INS \cite{Reznik05}).   Symmetry constraints and comparison of experimental results with the DFT calculations of peak positions and eigenvectors served as additional guides in peak assignments.  Because the detector banks extend out-of-plane, we were able to determine the c-axis modes as well (albeit with less precision). 

Figure \ref{fig:disp} shows that DFT calculations reproduce the phonon dispersions and intensities quite well. Although some small deviations from the calculated dispersions have been observed, they are on the order of 5\%, which is typical for the agreement between experiment and DFT calculations in general.  This good agreement has been achieved despite the fact that the calculations were carried out without inclusion of magnetism, although the structure was constrained to the experimental structure.  Our results for the Raman-active modes at the $\Gamma$-point are in good agreement with previous Raman studies \cite{Litvinchuk2008}.

We have not been able to establish the extent to which different branches mix with each other when they cross.  Such mixing should occur, but the resulting anti-crossing gaps were too small to be observed in our experiment.  The gaps could likely be determined by using a sample which is larger (for better statistics) and has a smaller mosaic (for better resolution).

Successful mapping of phonon dispersions allowed us to investigate temperature dependence of these phonon branches in order to look for phonon anomalies associated with magnetic ordering. Fig. \ref{fig:temp} shows that temperature-dependence of these phonons is weak and is consistent with small changes in lattice constants due to thermal expansion, i.e. these phonons are not particularly sensitive to the magnetic ordering transition.  We note some temperature variation does appear near points 4 and 8.  However, the polarization in those regions is primarily along the c-axis, and the weak intensity produces larger errors.  Along with the narrower range of data aquisition for T~= 220~K and 320~K, the uncertainty at these points are large enough to account for the variation seen here.

There is no measurable effect of the onset of long range magnetic order on any phonon branch measured so far as a part of this study (Fig. \ref{fig:temp}) or in previous studies, with the exception of transverse acoustic phonons whose eigenvectors correspond directly to the associated structural deformation \cite{Howard2012}.  It is also possible that phonons couple to the nematic fluctuations found at higher energies \cite{Reznik09, Harringer2011} and do not feel the effects of the phase transition.  These fluctuations may vary as a function of doping \cite{Lumsden10}, and may in turn influence the phonon spectrum.  The effect of doping on these high-energy phonons can now be investigated using the multizone approach.

One advantage of MPR can be the identification of enhanced electron-phonon coupling at wavevectors away from high symmetry directions that previously would have remained unobserved. For example, such anomalies have been found in Cr \cite{Lamago2010}.  We examined temperature-dependence of phonons away from high-symmetry directions in \SFA~and found that there are no anomalies there either.

\section{Conclusions}
We developed a multizone phonon refinement (MPR) technique for using time-of-flight neutron chopper spectrometers to extract phonon dispersions even when the phonon frequencies are spaced much more closely than the experimental resolution. It is based on fitting scattering intensities in multiple BZ. We used this technique to reliably measure closely-spaced phonons above 23~meV in ferropnictide superconductors. We assigned eigenvectors to these phonons and showed that they do not exhibit any unconventional temperature-dependence related to the magnetic ordering transition.  Our results pave the way for future detailed comparison of experimental and calculated phonon frequencies in iron-based superconductors and for future studies of materials with large unit cells where closely spaced phonon branches are expected, such as thermoelectrics and complex oxides.

\begin{acknowledgments}
D.P and D.R. were supported by the DOE, Office of Basic Energy Sciences, Office of Science, under Contract No. DE-SC0006939.  The research at ORNL's Spallation Neutron Source was sponsored by the Scientific User Facilities Division, Office of Basic Energy Sciences, U.S. Department of Energy.
\end{acknowledgments}

%
\end{document}